\documentclass[iop]{emulateapj_jw}

\usepackage{hyperref}
\usepackage{epsfig}
\usepackage{natbib}
\usepackage{graphicx}                
\usepackage{lscape}
\usepackage{verbatim}              
\usepackage{amsmath,amsthm,amsfonts,amsopn,amssymb} 
\usepackage{longtable}
\usepackage{pdflscape}
\usepackage{afterpage}
\usepackage{rotating}					
\usepackage{ulem}
\usepackage{epstopdf}  
\usepackage{multirow}
\usepackage{color}
\usepackage[left]{lineno}
\newcommand{\halfcheck}{\checkmark\kern-1.6ex\raisebox{.3ex}{\rotatebox[origin=c]{125}{--}}}

\received{}
\accepted{}

\slugcomment{to appear in ApJ}
\shorttitle{HJ Occurrence Rate}
\shortauthors{Wang et al.}

\begin{document}
\title{On the Occurrence Rate of Hot Jupiters in Different Stellar Environments}
\author{
Ji Wang\altaffilmark{1},
Debra A. Fischer\altaffilmark{1},
Elliott P. Horch\altaffilmark{2} and
Xu Huang\altaffilmark{3}
} 
\email{ji.wang@yale.edu}
%
\altaffiltext{1}{Department of Astronomy, Yale University, New Haven, CT 06511 USA}
\altaffiltext{2}{Department of Physics, Southern Connecticut State University, 501 Crescent Street, New Haven, CT 06515, USA}
\altaffiltext{3}{Department of Astrophysical Sciences, Peyton Hall, 4 Ivy Lane, Princeton University, Princeton, NJ 08540 USA}

\begin{abstract}

Many Hot Jupiters (HJs) are detected by the Doppler and the transit techniques. From surveys using these two techniques, however, the measured HJ occurrence rates differ by a factor of two or more. Using the California Planet Survey sample and the Kepler sample, we investigate the causes for the difference of HJ occurrence rate. First, we find that $12.8\%\pm0.24\%$ of HJs  are misidentified in the Kepler mission because of photometric dilution and subgiant contamination. Second, we explore the differences between the Doppler sample and the Kepler sample that can account for the different HJ occurrence rate. Third, we discuss how to measure the fundamental HJ occurrence rates by synthesizing the results from the Doppler and Kepler surveys. The fundamental HJ occurrence rates are a measure of HJ occurrence rate as a function of stellar multiplicity and evolutionary stage, e.g., the HJ occurrence rate for single and multiple stars or for main sequence and subgiant stars. While we find qualitative evidence that HJs occur less frequently in subgiants and multiple stellar systems, we conclude that our current knowledge of stellar properties and stellar multiplicity rate is too limited for us to reach any quantitative result for the fundamental HJ occurrence rates. This concern extends to $\eta_{\rm{Earth}}$, the occurrence rate of Earth-like planets. 


\end{abstract}


\section{Introduction}

Hot Jupiters (HJs) are among the most prominent astronomical discoveries of the past century~\citep{Mayor1995,Marcy1996}. Their existence challenged the previously accepted classic planet formation model, which is tailored for the solar system~\citep{Lissauer1993}. Our knowledge of planet formation has moved beyond the solar system ever since the first HJ discovery.

Among all exoplanets, HJs are the easiest to detect by the Doppler technique because of their short periods (P $<$ 10 days) and large planetary masses ($m>$ 0.1 $M_J$). Doppler planet surveys show that the occurrence rate of HJs is $\sim$1.0\%, although the exact numbers slightly differ for different surveys. For example, ~\citet{Wright2012} estimated that the occurrence rate of HJs is $1.2\%\pm0.38\%$ based on data obtained from the Keck and Lick Observatories. Other studies based on a similar set of data found the HJ occurrence rate to be $1.2\%\pm0.1\%$~\citep{Marcy2005} and $1.5\%\pm0.6\%$~\citep{Cumming2008}. Based on RV data obtained from the HARPS and ELODIE,~\citet{Mayor2011} measured the HJ occurrence rate at $0.89\%\pm0.36\%$.

The HJ occurrence rate measured from the Doppler planet surveys is at odds with measurements from transit surveys. In the Optical Galactic Lensing Experiment (OGLE) III survey, ~\citet{Gould2006} estimated that the HJ occurrence rate is  $0.31^{+0.43}_{-0.18}\%$. The number from the SuperLupus Survey is $0.10^{+0.27}_{-0.08}\%$~\citep{Bayliss2011}. {{A more recent result from the Kepler mission suggested that the HJ occurrence rate is $0.4\%\pm0.1\%$~\citep[$8.0\ R_\oplus\le R_P\le 32.0\ R_\oplus$, P $<$ 10 days,][]{Howard2012} or $0.43\%\pm0.05\%$~\citep[$6.0\ R_\oplus\le R_P\le 22.0\ R_\oplus$, 0.8 days $\le \rm{P} \le$ 10.0 days,][]{Fressin2013}. These numbers are a factor 2-3 smaller than those from Doppler surveys.}}

Potential explanations of the difference include stellar metallicity, age, and population~\citep{Howard2012,Wright2012, Gould2006,Bayliss2011}. Stars with higher metallicity are more likely to host HJs~\citep{Gonzalez1997,Santos2001,Fischer2005, Sousa2011,Wang2013b}. For example, at least 25\% of stars with twice the solar metallicity are orbited by a giant planet, and this number drops to less than 5\% for stars with solar metallicity~\citep{Udry2007}. In the sub-solar metallicity regime, the occurrence rate of HJ is found to be below 1\% at $0.37^{+0.6}_{-0.4}\%$~\citep{Mortier2012}. If the stellar metallicity for stars in Doppler surveys is systematically higher than those in transit surveys, then the HJ occurrence rate from the the Doppler surveys is higher than it is for the transit surveys. Given the strong dependence of the HJ occurrence rate on stellar metallicity~\citep{Fischer2005}, a metallicity difference of 0.15 dex may explain the difference of HJ occurrence rate. 

Stellar age may also explain the difference of HJ occurrence rate. If Kepler stars are on average older, then the fraction of evolved stars is higher. Since there is evidence that evolved stars host fewer HJs~\citep{Bowler2010,Johnson2010,Schlaufman2013}, a higher fraction of evolved stars may lead to a lower HJ occurrence rate for the Kepler sample. 

Another possible explanation may be stellar population. Transit surveys usually target galactic bulge~\citep[e.g.,][]{Gould2006,Bayliss2011}, where stellar population is dominated by M dwarfs~\citep{Henry1997}. In comparison, Doppler surveys usually selected solar-type stars~\citep[e.g.,][]{Valenti2005,Sousa2011}. {{Since HJs occur less frequently around M dwarfs~\citep{Johnson2010,Bonfils2013}, a sample of stars with higher fraction of M dwarfs leads to a lower HJ occurrence rate.}} However, the stellar population difference is less of a concern for the comparison between the Kepler sample and the Doppler sample. Unlike other transit surveys, Kepler stars mainly consist of solar-type stars~\citep{Brown2011}.

In this paper, we investigate other possibilities that may reconcile the difference of HJ occurrence rate from the Kepler and Doppler survey. In \S \ref{sec:simul}, we quantify the fraction of HJs that are misidentified from the Kepler mission because of photometric dilution and subgiant contamination. In \S \ref{sec:Method}, we summarize the potential explanations for the discrepancy of  HJ occurrence rate. In \S \ref{sec:fund} and \S \ref{sec:hj_occ}, we synthesize the Kepler and Doppler results and discuss how to probe the fundamental HJ occurrence rate as a function of stellar multiplicity rate and evolutionary stage. In \S \ref{sec:Summary}, we discuss our conclusions and their implications to future investigations of planet occurrence rate. 

\section{Fraction of Misidentified HJs}
\label{sec:simul}

\subsection{Simulated HJs}
\label{sec:hjs}

\begin{figure}
\begin{center}
\includegraphics[width=16cm,height=10.5cm]{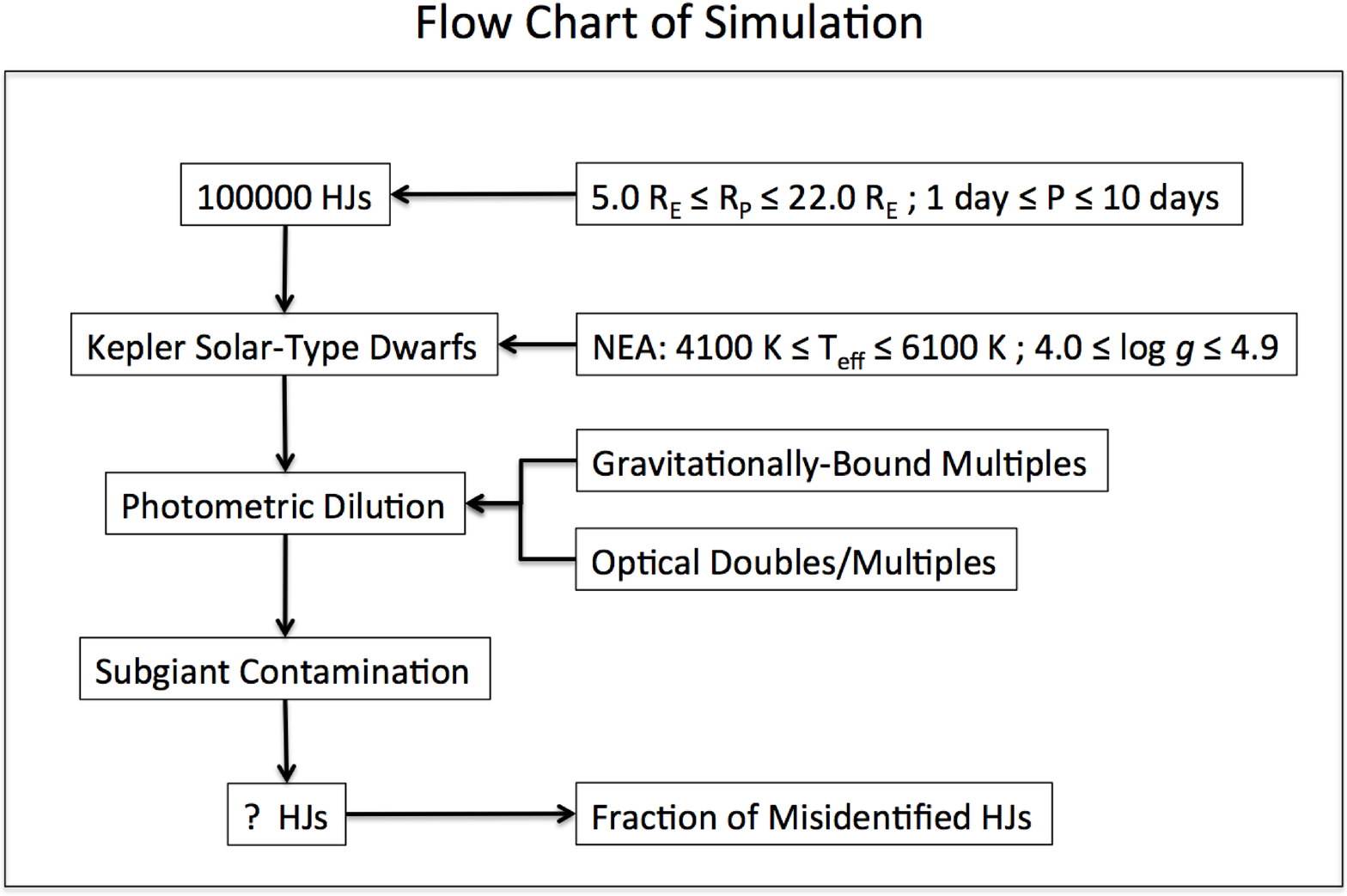} \caption{Flow chart of simulation to quantify the fraction of HJs that are misidentified as smaller planets due to two effects: photometric dilution and subgiant contamination.
\label{fig:flowchart}}
\end{center}
\end{figure}

We attempt to quantify the fraction of HJs that are misidentified as smaller planets due to two effects: photometric dilution and subgiant contamination. Fig. \ref{fig:flowchart} shows the flow chart of our simulation. We start with 100,000 HJs ($1\ \rm{day}\le P \le10\ \rm{days}$, $5.0\ R_\oplus\le R_P\le 22.0\ R_\oplus$). The lower limit of HJs roughly corresponds to 0.1 $M_J$ assuming the mass-radius relationship used in~\citet{Lissauer2011}. 
The radius distribution of these HJs follows a power law of $-2.92\pm0.11$~\citep{Howard2012}. The period distribution between 2 and 10 days follows a power law of -1, i.e., uniform distribution in logarithmic space. The period distribution between 1 and 2 days follows the same power law distribution but with half the probability of $2\ \rm{day}\le P \le10\ \rm{days}$. This assumption accounts for the lack of HJs within the 2-day orbital period~\citep{Howard2012}.

\subsection{Kepler Solar-Type Dwarf Stars}
\label{sec:keplerstar}

We obtain Kepler stellar properties from the NASA Exoplanet Archive~\citep[NEA,][]{Huber2014}\footnote{http://exoplanetarchive.ipac.caltech.edu/}. We apply a cut in effective temperature ($4100\ K\le T_{\rm{eff}}\le 6100\ K$) and surface gravity ($4.0\le\log g\le4.9$), which results in 109,335 solar-type dwarf stars in the Kepler sample. The cut is consistent with the solar-type star definition from~\citet{Howard2012} for a proper comparison to previous result. 

\subsection{Photometric Dilution in Multiple Stellar Systems}
\label{sec:multiplestar}

The transit signal of a planet is diluted by the flux of companion stars. The dilution leads to two consequences: (1), a missing planet or (2), a planet with an underestimated planet radius in the detection case. For the Kepler mission, a planet is missed when the detection signal to noise ratio (SNR) is lower than 7.1~\citep{Jenkins2010}. The SNR is calculated based on the following equation~\citep{Wang2014b}:
\begin{equation}
\label{eq:snr}
S/N=\frac{\delta}{CDPP_{\rm{eff}}}\sqrt{N_{\rm{transits}}},
\end{equation}
where $\delta$ is transit depth, $CDPP_{\rm{eff}}$ is the effective combined differential photometric precision~\citep{Jenkins2010b}, a measure of photometric noise, and $N_{\rm{transits}}$ is the number of observed transits. The transit depth is calculated by the following equation:
\begin{equation}
\label{eq:delta}
\delta=\frac{R^2_{\rm{PL}}}{R^2_{\ast}}\frac{F_\ast}{F_\ast+F_c},
\end{equation}
where $R_{\rm{PL}}$ is planet radius, $R_\ast$ is the radius of the star that the planet is transiting, $F$ denotes flux, and subscripts $\ast$ and $c$ indicate the planet host star and the contaminating star, respectively. Throughout the paper, we only consider the case in which a HJ transits the primary star because we focus on HJs around solar-type stars and the secondary star may not be a solar-type star in most cases. 

Even if a planet is detected in the presence of photometric dilution, the photometric dilution will cause some HJs to be misidentified as smaller planets. These misidentified HJs are not accounted in the statistics for HJs. Thus, the fraction of misidentified HJs needs to be quantified and corrected for. The measured planet radius ($R^\prime_{\rm{PL}}$) can be calculated by the following equation:
\begin{equation}
\label{eq:delt}
R^\prime_{\rm{PL}}=R_{\rm{PL}}\sqrt{\frac{F_\ast}{F_\ast+F_c}}.
\end{equation}

\subsubsection{Gravitationally-Bound Multiple Stellar Systems}
\label{sec:bound}

Companions stars may be gravitationally bound or optical doubles or multiples (i.e., unbound). We consider these two cases separately. For gravitationally bound multiple stellar systems, the stellar multiplicity rate for planet host stars~\citep{Wang2014a,Wang2014b} is significantly lower than that for the solar neighborhood~\citep{Raghavan2010}. We adopt the stellar multiplicity rate measured from~\citet{Wang2014b} for stellar separations smaller than 1000 AU, i.e., 24\%$\pm$7\%. Beyond 1000 AU, we adopt the log-normal distribution of stellar separation from~\citet{Duquennoy1991}. At these separations, there is no significant difference of stellar multiplicity between ~\citet{Wang2014b} and~\citet{Duquennoy1991}. In addition, the result from~\citet{Wang2014b} beyond 1000 AU is prone to contamination of optical doubles and multiples, which will be addressed in the following section. In our simulation for gravitationally-bound systems, mass ratio of stellar companions to the primary star follows a normal distribution with mean and standard deviation of 0.23 and 0.42~\citep{Duquennoy1991}. The conversion between stellar mass and radius is based on~\citet{Feiden2012}. The conversion from stellar mass to stellar flux in the Kepler band is based on~\citet{Kraus2007}.


\subsubsection{Optical doubles and multiples}
\label{sec:unbound}

For photometric dilution caused by optical doubles and multiples, we use the TRILEGAL galaxy model~\citep{Girardi2005} to study the probability of such cases.~\citet{Horch2014} describes the process in detail. Ten fields with a field of view of 1 square degree are simulated. These fields have different galactic latitudes so the combination of the results from the fields gives a better statistical result of the entire Kepler field of view. In each of field, TRILEGAL model is used to construct a simulated stellar population with effective temperature distribution that is similar to Kepler stars. Binary parameters are turned off in the simulation because we focus on optical doubles and multiples, gravitationally bound multiple stellar systems have already been discussed.

\begin{figure}
\begin{center}
\includegraphics[width=10cm,height=16cm]{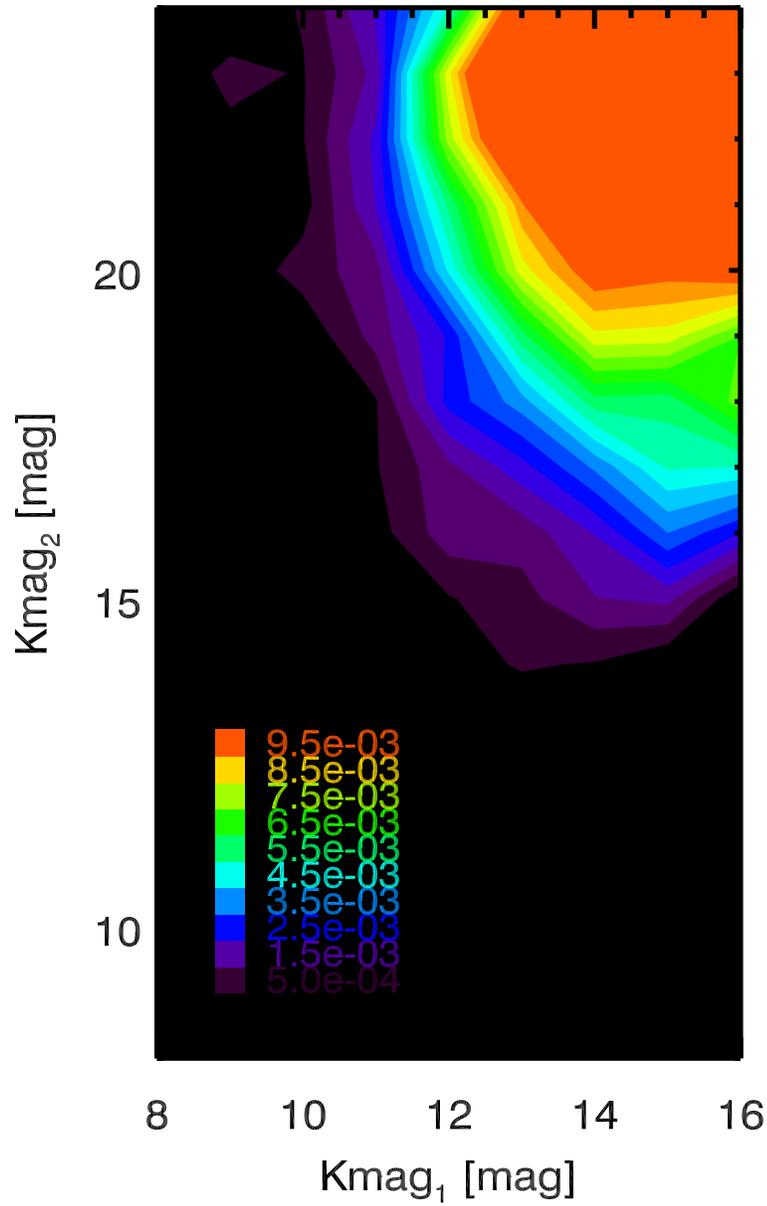} \caption{Contours of joint probability distribution of Kmag$_{1}$ and Kmag$_{2}$ for optical doubles and multiples in the Kepler field of view. Kmag$_{1}$ is the Kepler magnitude of the primary star and Kmag$_{2}$ is the Kepler magnitude of the secondary star. 55.7\% of Kepler stars have at least one visual unbound stellar companion. Fainter Kepler stars are more likely to have visual unbound stellar companions with small delta magnitudes. Therefore, planets around faint Kepler stars are more likely to be misidentified as smaller planets. 
\label{fig:fk1k2}}
\end{center}
\end{figure}

From the simulations, we calculated the probability of optical doubles or multiples. {{We find that 55.7\% of the simulated stars have at least one visual companion within 4$^{\prime\prime}$ down to Kepler magnitude of 27.0.}} The separation of 4$^{\prime\prime}$ corresponds to the pixel scale of the Kepler CCD detector. {{Since stellar companions with small differential magnitudes cause large adjustment of planet radius (Equation \ref{eq:delt}), we also report the fraction of simulated stars with stellar companions down to a certain differential magnitude. For differential magnitudes of 1, 3, and 5 mag, the fractions of simulated stars with stellar companions are 2.8\%, 11.3\%, and 26.7\%, respectively.}} Fig. \ref{fig:fk1k2} shows a contour plot of joint probability of Kmag$_{1}$ and Kmag$_{2}$, where Kmag$_{1}$ is the Kepler magnitude of the primary star and Kmag$_{2}$ is the Kepler magnitude of the secondary star. Optical doubles and multiples with small delta magnitude are more likely for faint Kepler stars (Kmag$_{1}$ $\geq$ 13.0) than for bright Kepler stars (Kmag$_{1}$ $\leq$ 13.0). In a transit observation, measured planet radius is affected more by brighter stellar companions (see Equation \ref{eq:delt}), i.e., secondary star with smaller delta magnitude, so misidentified HJs are more likely take place for faint Kepler stars. 

\subsection{Contamination of Subgiants}
\label{sec:subgiant}

In a transit observation, the ratio of planet radius to star radius is measured. When converting the ratio to planet radius, the star radius needs to be multiplied. A systematic error in star radius estimation can lead to inaccurate planet radius and thus misidentify a planet into an incorrect category. While we use the NEA stellar property catalog to select solar-type dwarf stars, chances are that some the selected stars are actually subgiant stars. Reported planet radii from the NEA may be systematically lower than they should be with the contamination of subgiants.~\citet{Bastien2014} used short-time scale photometric variation of bright Kepler stars (Kmag $\leq$ 13.0) to infer stellar surface gravity and radius. They found that 48\% of Kepler stars are modestly evolved subgiant stars. In comparison, 27\% of Kepler stars are subgiants from the NEA statistics. The comparison indicates that 29\% of Kepler "dwarf" stars according to the NEA are actually subgiants. These subgiants are on average 20\%-30\% larger than reported in the NEA. Therefore, radii of $\sim$30\% of Kepler planets are underestimated by 20\%-30\%. We use a value of 25\% for the percentage by which the stellar radius is underestimated. To account for subgiant contamination in the Kepler dwarf sample, we randomly assign 30\% of {{simulated stars}} with HJs as subgiants that are misidentified as "dwarf" stars. We adopt the adjusted subgiant radius to calculate SNR based on Equation \ref{eq:snr}. If the SNR is lower than 7.1, then the HJ is marked as a missing HJ due to insufficient SNR. In the case of detection ($\rm{SNR}\ge7.1$), the radius of the HJ is recorded with a smaller value by a factor of 1.25. 

\subsection{Result}
\label{sec:result}

\begin{figure}
\begin{center}
\includegraphics[width=16cm,height=15.0cm]{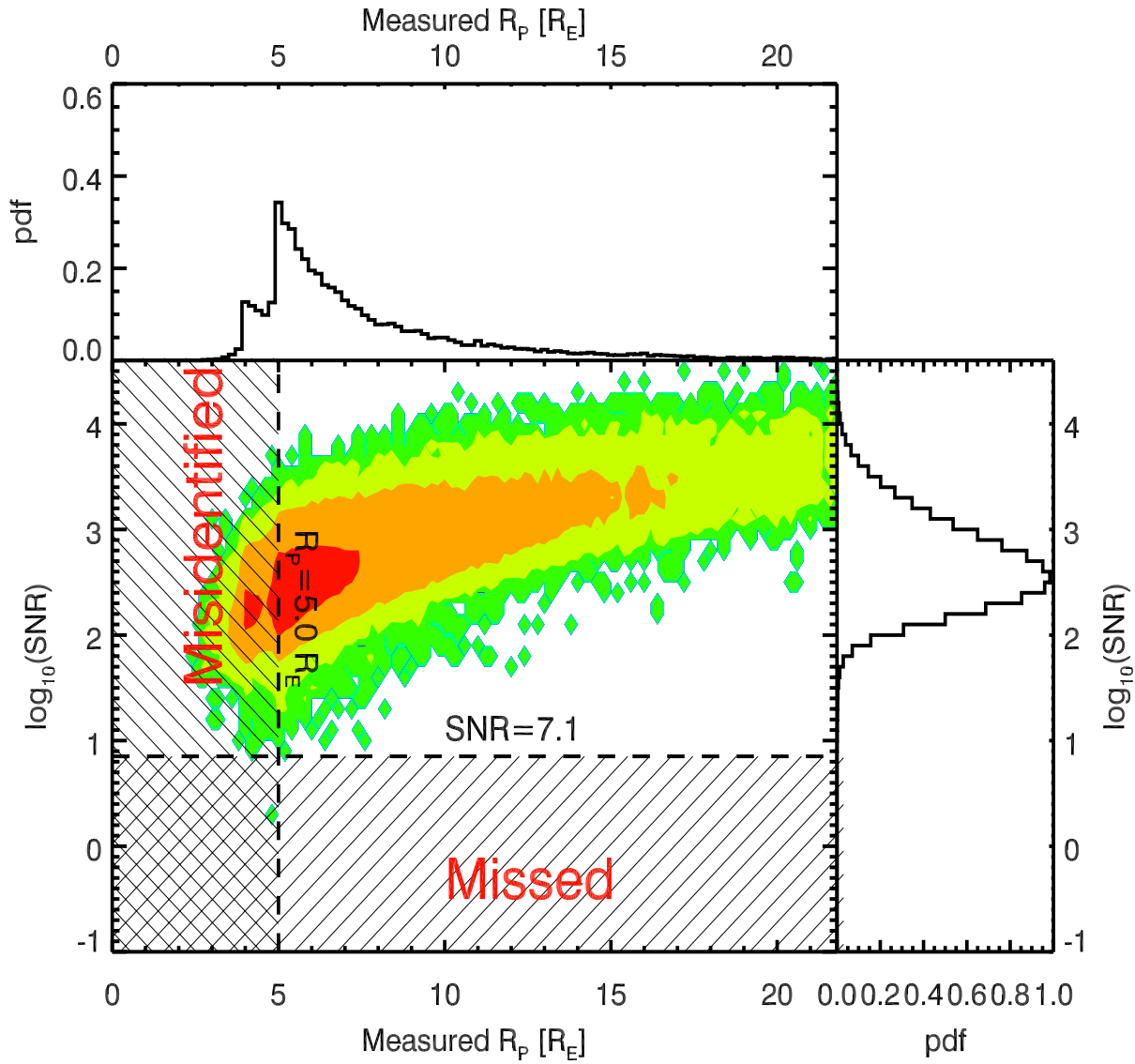} \caption{Result of simulation to quantify the fraction of HJs that misidentified because of photometric dilution and subgiant contamination. Color contours show the joint distribution of SNR and measured planet radius of simulated HJs. The marginalized distributions of SNR and measured planet radius are shown on the side of each axis. We find that 12.48\%$\pm$0.24\% of HJs are misidentified due to photometric dilution and subgiant contamination. 
\label{fig:HJ}}
\end{center}
\end{figure}

We repeat the above simulation for 100 times and calculate the fraction of HJs that are misidentified as smaller planets. We find that 12.48\%$\pm$0.24\% of HJs are misidentified due to photometric dilution and subgiant contamination. Specifically, 2.11\%$\pm$0.08\% and 1.17\%$\pm$0.06\% of misidentifications are caused by photometric dilution due to gravitationally bound multiple stellar systems and optical doubles/multiples, respectively. 9.59\%$\pm$0.15\% of misidentifications are caused by subgiant contamination. Therefore, subgiant contamination is the major cause that is responsible for HJ misidentification. Fig. \ref{fig:HJ} shows a contour plot for distribution of HJ planet radius and SNR after considering the effects of photometric dilution and subgiant contamination. Because of large signal of HJs, they are rarely missed due to an insufficient SNR. The fraction of missing HJs is always less than 0.005\% and is thus negligible compared to misidentified HJs. 

\section{Differences Between the Doppler and Kepler Sample}
\label{sec:Method}

{{While there are numerous works on the HJ occurrence rate for the Doppler sample and the Kepler sample, we focus on two works that have the most consistency between samples, i.e., ~\citet{Wright2012} and ~\citet{Howard2012}.}}{{ ~\citet{Howard2012} used the Kepler sample to measure the occurrence rate of HJs, $f_{\rm{Kepler}}$. Interpolating their result, the occurrence rate is 0.60$\pm$0.10\% for HJs with $5.0\ R_\oplus\le R_P\le 22.0\ R_\oplus$ and P $<$ 10 days.}} Emulating the Howard et al. sample, but using the California Planet Survey (CPS) sample, ~\citet{Wright2012} found that the occurrence rate of HJs for the Doppler sample is 1.20$\pm$0.38\%. Although marginally significant, ~\citet{Wright2012} speculated that the difference can be accounted for by metallicity and evolutionary stage difference between the Doppler and Kepler sample.

\subsection{Metallicity Difference}
\label{sec:metallicity}

The occurrence rate of HJs is a strong function of stellar metallicity~\citep{Gonzalez1997,Santos2001,Santos2004,Fischer2005,Johnson2010,Sousa2011,Wang2013b}. A slight difference of stellar metallicity between the Doppler and Kepler sample will result in a significant difference of HJ occurrence rate. The CPS survey targets stars in the solar neighborhood. The metallicity distribution of these nearby stars can be obtained from the SPOCS (Spectroscopic Properties Of Cool Stars) catalog~\citep{Valenti2005}. We apply the effective temperature and surface gravity cut ($4100\ K\le T_{\rm{eff}}\le 6100\ K$, $4.0\le\log g\le4.9$) to the SPOCS catalog to select solar-type stars. The cut is consistent with the solar-type star definition from~\citet{Howard2012} for a proper comparison to the Kepler sample. A total of 694 stars from the SPOCS are selected. The mean and median metallicity for the Doppler sample is -0.01 and 0.02 dex. In comparison, the mean and median metallicity of the Kepler sample -0.04 and -0.03 dex for 12,400 Kepler stars~\citep{Dong2014}. The metallicity difference between the Doppler sample and the Kepler sample is thus 0.03 dex (mean) or 0.05 dex (median), which results in a factor of 1.15 (mean) or 1.26 (median) difference in the HJ occurrence rate if using a power law of 2.0~\citep{Fischer2005}. Therefore, the stellar metallicity alone cannot account for the difference of HJ occurrence rate. {{Given the sample size for metallicity determination, i.e., 694 for the Doppler sample and 12,400 for the Kepler sample, the standard error of metallicity measurement is small. However, the systematic error of metallicity measurement is estimated at $\sim$0.05 dex~\citep{Hinkel2014}, which is much higher than the standard error. Taking the systematic error into account, stellar metallicity may lead to a factor of 1.58 difference, but still not adequate to explain the factor of 2-3 difference of HJ occurrence rate.}}

\subsection{Evolutionary Stage Difference}
\label{sec:evo}

HJs occur less frequently around evolved stars than around main sequence stars~\citep{Johnson2010,Bowler2010}. Inclusion of evolved stars may lower the HJ occurrence rate for a survey. The CPS uses SPOCS as the input catalog. Following the selection criteria of~\citet{Wright2012}, $V<8$, $B-V<1.2$, and $\Delta M_V<2.5$ mag. The latter requirement corresponds roughly to log $g > 3.5$. There are 745 stars in SPOCS meeting the criteria. Among these stars, 12.6\% of stars have log $g<4.1$. The range $3.5<$log $g<4.1$ is what is defined as subgiants in~\citet{Bastien2014} for the Kepler sample. {{They found that 48\% of bright Kepler stars are subgiants. Therefore, the fraction of subgiants is much higher for the Kepler sample than for the Doppler sample.}} The difference of subgiant fraction can leads to the difference of HJ occurrence rate. 

\subsection{Stellar Multiplicity Rate Difference}
\label{sec:multi}

Doppler planet surveys usually target single stars and stars without nearby (sep $<2^{\prime\prime}$) stellar companions~\citep{Wright2012}, so the stellar multiplicity rate for the Doppler sample should be lower than what is known for stars in the solar neighborhood~\citep{Duquennoy1991,Raghavan2010}. Given the median distance of 30 pc for the stars selected in~\citet{Wright2012}, a separation of 2$^{\prime\prime}$ corresponds to 60 AU, which is roughly the peak of stellar separation distribution~\citep{Raghavan2010}. Therefore, the stellar multiplicity rate for the Doppler sample is at most half of the stellar multiplicity rate in the solar neighborhood, i.e., $\sim$23\%. In comparison, stars in the Kepler sample are in general much further away at $\sim$200-1000 pc. There is no strong selection bias against stars with stellar companions~\citep{Brown2011}. Kepler pixel scale is 4.0$^{\prime\prime}$  and photometric aperture is usually at least twice as large, so the Kepler mission observes both single stars and multiple stellar systems with stellar separations almost extends to the tail of stellar separation distribution. Therefore, the stellar multiplicity rate of the Kepler sample should be higher than the Doppler sample. Given that planet formation is suppressed in multiple star systems~\citep[e.g.,][]{Wang2014b}, including more multiple star systems in the sample may lead to a lower planet occurrence rate. 

\subsection{False Positive Rate for HJs from Kepler}
\label{sec:fap}

Not all HJ candidates detected by the Kepler are bona fide planets. The false positive rate of HJs is estimated between 10\% and 35\%~\citep{Morton2011,Santerne2012,Fressin2013}. Considering the false positive rate for HJs, the HJ occurrence rate from the Kepler mission should be even lower.~\citet{Fressin2013} found the HJ occurrence rate to be 0.43\% accounting for the false positive rate. 

\section{Probing the Fundamental HJ Occurrence Rate}
\label{sec:fund}

We have covered a variety of potential causes that may account for the difference of HJ occurrence rates from the Doppler and Kepler sample. The following equations summarize how these potential causes can be combined to probe the fundamental HJ occurrence rate as a function of stellar multiplicity and evolutionary stage. The measured HJ occurrence rate from the Kepler or the Doppler survey is a combined result of fundamental HJ occurrence rates:
\begin{equation} 
\label{eq:f_hj_sgr}
f=f_{\rm{MS}}\times(1-\rm{SGR})+f_{\rm{SG}}\times\rm{SGR}, \rm{or}
\end{equation}
\begin{equation} 
\label{eq:f_hj_mr}
f=f_{\rm{single}}\times(1-\rm{MR})+f_{\rm{multiple}}\times\rm{MR},
\end{equation}
Where $f$ is measured HJ occurrence rate, MS represents main sequence, SG represents subgiant, MR is stellar multiplicity rate and SGR is the fraction of subgiants in the sample. We have two measurements from the Kepler and Doppler surveys. However, there are four fundamental HJ occurrence rates that we wish to solve for, $f_{\rm{MS},\rm{single}}$, $f_{\rm{MS},\rm{multiple}}$, $f_{\rm{SG},\rm{single}}$, and $f_{\rm{SG},\rm{multiple}}$. In addition, there are correcting factors for the difference between the Kepler and Doppler sample. For example, we should consider $\xi_{[\rm{Fe/H}]}$, the correction factor for the metallicity difference (\S \ref{sec:metallicity}), $\xi_{\rm{misidentified}}$,  the correction factor for the misidentified HJs due to photometric dilution and subgiant contamination (\S \ref{sec:result}), and $\rm{FPR}$, the false positive rate for HJs from the Kepler sample (\S \ref{sec:fap}). 

We wish to solve for these fundamental HJ occurrence rates. However, current knowledge of the Kepler and Doppler sample is inadequate. Table \ref{tab:current} summarizes our current knowledge of the parameters in above equations. Different Doppler planet surveys converge to $f_{\rm{Doppler}}\sim1\%$ for HJ occurrence rate~\citep{Mayor2011,Cumming2008, Wright2012}. {{$f_{\rm{Kepler}}$ measured from different independent works agree at $\sim0.4-0.6\%$~\citep{Howard2012,Fressin2013}. }}The correcting factor $\xi_{[\rm{Fe/H}]}$ for the metallicity difference is still uncertain to $\sim$0.05 dex which is the systematic error of abundance analysis~\citep{Hinkel2014}. The 0.05 dex difference would result in 25\% change in HJ occurrence rate. We quantify $\xi_{\rm{misidentified}}$ in this work, but its value depends on assumptions of other parameters, such as MR and SGR. FPR for the Kepler sample has been studied extensively, but its value is still not well constrained, ranging from 10\% to 35\%~\citep{Morton2011,Santerne2012,Fressin2013}. 

Understanding the stellar properties and statistics is crucial in solving Equation \ref{eq:f_hj_sgr} and \ref{eq:f_hj_mr}. Determination of SGR requires measurement of log $g$. While the stars in the Doppler sample have relatively well-determined log $g$~\citep{Valenti2005,Sousa2011}, log $g$ distribution for the Kepler sample is still uncertain especially for the faint end~\citep{Bastien2014}. MR$_{\rm{MS}}$ for the Doppler sample is not well constrained because of selection bias. While there is no strong selection bias against multiple star systems for the Kepler sample, but it is not known whether MR$_{\rm{MS}}$ is the same for the Kepler sample as for the solar neiborhood~\citep{Raghavan2010}. The knowledge of MR$_{\rm{SG}}$ is more uncertain for both the Kepler and Doppler sample. 

\section{HJ Occurrence Rate in the Solar Neighborhood}
\label{sec:hj_occ}

It is very challenging, if not impossible, to probe the fundamental HJ occurrence rates given the uncertainties of the parameters required for the calculation. In addition, we need some additional constraints to solve for the four variables, $f_{\rm{MS},\rm{single}}$, $f_{\rm{MS},\rm{multiple}}$, $f_{\rm{SG},\rm{single}}$, and $f_{\rm{SG},\rm{multiple}}$, with only two measurements from the Doppler and Kepler sample. However, the fundamental HJ occurrence rates can be calculated under two extreme and yet unlikely circumstances where the four variables are reduced to two. While the following two extreme cases do not necessarily reflect the reality, but they help us to understand how we can use Equation \ref{eq:f_hj_sgr} and \ref{eq:f_hj_mr} to probe the fundamental HJ occurrence rates. 

\subsection{Extreme Case I: $f_{\rm{single}}=f_{\rm{multiple}}$}
\label{sec:hj_occ_sm}

It is still under debate how the HJ occurrence rate in single stars compares to that for multiple stellar systems. On one hand, evidence of suppressed planet formation is found in multiple stellar systems~\citep[e.g.,][]{Wang2014a,Wang2014b}. On the other hand, a stellar companion may facilitate the formation of HJs via Kozai perturbation~\citep[e.g.,][]{Wu2003}, although a recent study shows that HJ formation via this channel may have an upper limit of 44\%~\citep{Dawson2012b}. Therefore, it is not an unreasonable assumption that HJ occurrence rate is the same in single stars as in multiple stellar systems. 

With this assumption, Equation \ref{eq:f_hj_sgr} and \ref{eq:f_hj_mr} are reduced to Equation \ref{eq:f_hj_sgr} with two variables, $f_{\rm{MS}}$ and $f_{\rm{SG}}$. Two measurements of $f$ are available. One is from the Kepler sample and the other one is from the Doppler sample. In this calculation, we adopt $f_{\rm{Kepler}}=0.6\%\pm0.1\%$~\citep{Howard2012}, $f_{\rm{Doppler}}=1.20\%\pm0.38\%$~\citep{Wright2012}. We apply correction factors $\xi_{\rm{[Fe/H]}}=1.26$, $\xi_{\rm{misidentified}}=1.14$, and $\rm{FAP}=17\%$~\citep{Fressin2013}. We adopt SGR for the Kepler sample to be 48\%~\citep{Bastien2014}, and 12.6\% for the Doppler sample~\citep{Valenti2005}. Substituting all the adopted values into Equation \ref{eq:f_hj_sgr} for the Kepler and Doppler sample, we infer that $f_{\rm{MS}}=1.37\%\pm0.52\%$ and $f_{\rm{SG}}=0.00\%\pm0.61\%$. The uncertainties for $f_{\rm{MS}}$ and $f_{\rm{SG}}$ are based only on the uncertainties of $f_{\rm{Kepler}}$ and $f_{\rm{Doppler}}$. The result indicates that HJs are much less common around subgiant stars than around main sequence stars, which is consistent with observations~\citep{Bowler2010,Johnson2010}. 

\subsection{Extreme Case II: $f_{\rm{MS}}=f_{\rm{SG}}$}
\label{sec:hj_occ_ms}

On the other hand, Doppler observations of subgiants have not completely ruled out the possibility that HJs may be as common around subgiants as around main sequence stars. The sample in ~\citet{Bowler2010} contains 31 subgiants. It is difficult to rule out that $f_{\rm{MS}}\ne f_{\rm{SG}}$ with the small sample size. If we assume that $f_{\rm{MS}}=f_{\rm{SG}}$, Equation \ref{eq:f_hj_sgr} and \ref{eq:f_hj_mr} are reduced to Equation \ref{eq:f_hj_mr} with two variables, $f_{\rm{single}}$ and $f_{\rm{multiple}}$. In this case, we adopt $\rm{MR}=46\%$ for the Kepler sample, which is the same as the stellar multiplicity rate in the solar neighborhood~\citep{Raghavan2010}. We adopt $\rm{MR}=5\%$, which is the stellar multiplicity rate for all the known planets detected by the Doppler technique~\citep{Wright2011}. Adopting the same values for other parameters as the previous case, we infer that $f_{\rm{single}}=1.26\%\pm0.43\%$ and $f_{\rm{multiple}}=0.08\%\pm0.55\%$. The uncertainties for $f_{\rm{single}}$ and $f_{\rm{multiple}}$ are based only on the uncertainties of $f_{\rm{Kepler}}$ and $f_{\rm{Doppler}}$.

We emphasize that the two extreme cases do not necessarily reflect the reality. {{In fact, one extreme assumption leads to a result that is significantly different from the other extreme assumption. Therefore, a meaningful solution of these fundamental HJ occurrence rate must lie in between these two extremes.}} To better understand the fundamental HJ occurrence rate, we need to have better knowledge of the stellar properties of the Kepler and Doppler sample and some other external constraints. 

\section{Summary and Discussion}
\label{sec:Summary}

\subsection{Summary}

We conduct simulation to investigate the fraction of HJs missed or misidentified by the Kepler mission due to the effects of photometric dilution and subgiant contamination. Despite these two effects, the Kepler mission rarely missed any HJs because of their large photometric signal. However, 12.48\%$\pm$0.24\% of HJs are misidentified as smaller planet due to photometric dilution and subgiant contamination. Therefore, the occurrence rate of HJs from the Kepler mission should be revised upward by $\sim$14\%. 

We review the differences of stellar properties and stellar multiplicity between the Kepler and Doppler sample. The differences, together with the misidentified HJs, can potentially explain the discrepancy of HJ occurrence rate. We discuss the fundamental HJ occurrence rate as a function of stellar multiplicity and evolutionary stage, i.e., HJ occurrence rate for single and multiple stellar systems and for main sequence and subgiant stars. We attempt to solve for the fundamental HJ occurrence rate by synthesizing the results from the Kepler and Doppler sample. However, our current knowledge of stellar properties and stellar multiplicity rate is too limited for us to draw any quantitative conclusion. In two special cases, though not entirely realistic, we find qualitative evidence that HJs occur less frequently around subgiant stars and in multiple stellar systems. 

\subsection{Future Work on the Fundamental HJ Occurrence Rates}

To quantitatively calculate the fundamental HJ occurrence rate, $f_{\rm{MS},\rm{single}}$, $f_{\rm{MS},\rm{multiple}}$, $f_{\rm{SG},\rm{single}}$, and $f_{\rm{SG},\rm{multiple}}$, we must need external constraints in addition to the Doppler and Kepler results. For example, one can calculate $f_{\rm{MS},\rm{single}}$ by carefully selecting main sequence single stars in the Doppler survey. {{Alternatively, one can select main sequence stars and sub giants in the Kepler sample based on the "Flicker" method~\citep{Bastien2014} or spectroscopic methods. From these two sample of Kepler stars, one can calculate the ratio of  $f_{\rm{MS}}$ and  $f_{\rm{SG}}$ assuming similar stellar multiplicity rate for main sequence and sub giant stars.}} In addition, the ratio of $f_{\rm{MS},\rm{single}}$ to $f_{\rm{MS},\rm{multiple}}$ is provided by~\citet{Wang2014b}, although the caveat is that the ratio is for a sample of planet candidates that are mostly small planets. With more constraints, the fundamental planet occurrence may be calculated quantitatively. However, much more effort is needed to make sure the calculation is meaningful. Future missions such as K2~\citep{Howell2014} and TESS~\citep{Ricker2014} may provide independent measurement of HJ occurrence rate, i.e., $f_{\rm{K2}}$ and $f_{\rm{TESS}}$, these results can be incorporated to calculate the fundamental HJ occurrence rate given that stellar properties and stellar multiplicity rate are known to a certain accuracy. 

\subsection{Implications to $\eta_{\rm{Earth}}$}

The findings in this paper can be used for the determination of $\eta_{\rm{Earth}}$, the occurrence rate of an Earth-like planet in the habitable zone. This issue has already been complicated by the definition of the habitable zone~\citep{Seager2013,Kopparapu2013}. Furthermore, the determination of $\eta_{\rm{Earth}}$ should take into account the photometric dilution and subgiant contamination. Finally, the measurement of $\eta_{\rm{Earth}}$ from either the Kepler or the Doppler survey is a combination of the fundamental $\eta_{\rm{Earth}}$ as a function of stellar multiplicity rate and evolutionary stage (Equation \ref{eq:f_hj_sgr} and \ref{eq:f_hj_mr}). These fundamental $\eta_{\rm{Earth}}$ can be precisely determined only if we have a good understanding of the stellar properties and stellar multiplicity rate of the sample.

\noindent{\it Acknowledgements} We thank the referee Ronald Gilliland for his insightful comments and suggestions which substantially improve the paper. This research has made use of the NASA Exoplanet Archive, which is operated by the California Institute of Technology, under contract with the National Aeronautics and Space Administration under the Exoplanet Exploration Program.


\begin{thebibliography}{50}
\expandafter\ifx\csname natexlab\endcsname\relax\def\natexlab#1{#1}\fi

\bibitem[{{Bastien} {et~al.}(2014){Bastien}, {Stassun}, \&
  {Pepper}}]{Bastien2014}
{Bastien}, F.~A., {Stassun}, K.~G., \& {Pepper}, J. 2014, \apjl, 788, L9

\bibitem[{{Bayliss} \& {Sackett}(2011)}]{Bayliss2011}
{Bayliss}, D.~D.~R., \& {Sackett}, P.~D. 2011, \apj, 743, 103

\bibitem[{{Bonfils} {et~al.}(2013){Bonfils}, {Delfosse}, {Udry}, {Forveille},
  {Mayor}, {Perrier}, {Bouchy}, {Gillon}, {Lovis}, {Pepe}, {Queloz}, {Santos},
  {S{\'e}gransan}, \& {Bertaux}}]{Bonfils2013}
{Bonfils}, X., {et~al.} 2013, \aap, 549, A109

\bibitem[{{Bowler} {et~al.}(2010){Bowler}, {Johnson}, {Marcy}, {Henry}, {Peek},
  {Fischer}, {Clubb}, {Liu}, {Reffert}, {Schwab}, \& {Lowe}}]{Bowler2010}
{Bowler}, B.~P., {et~al.} 2010, \apj, 709, 396

\bibitem[{{Brown} {et~al.}(2011){Brown}, {Latham}, {Everett}, \&
  {Esquerdo}}]{Brown2011}
{Brown}, T.~M., {Latham}, D.~W., {Everett}, M.~E., \& {Esquerdo}, G.~A. 2011,
  \aj, 142, 112

\bibitem[{{Cumming} {et~al.}(2008){Cumming}, {Butler}, {Marcy}, {Vogt},
  {Wright}, \& {Fischer}}]{Cumming2008}
{Cumming}, A., {Butler}, R.~P., {Marcy}, G.~W., {Vogt}, S.~S., {Wright}, J.~T.,
  \& {Fischer}, D.~A. 2008, \pasp, 120, 531

\bibitem[{{Dawson} {et~al.}(2012){Dawson}, {Murray-Clay}, \&
  {Johnson}}]{Dawson2012b}
{Dawson}, R.~I., {Murray-Clay}, R.~A., \& {Johnson}, J.~A. 2012, ArXiv e-prints

\bibitem[{{Dong} {et~al.}(2014){Dong}, {Zheng}, {Zhu}, {De Cat}, {Fu}, {Yang},
  {Zhang}, {Jin}, \& {Zhang}}]{Dong2014}
{Dong}, S., {et~al.} 2014, \apjl, 789, L3

\bibitem[{{Duquennoy} \& {Mayor}(1991)}]{Duquennoy1991}
{Duquennoy}, A., \& {Mayor}, M. 1991, \aap, 248, 485

\bibitem[{{Feiden} \& {Chaboyer}(2012)}]{Feiden2012}
{Feiden}, G.~A., \& {Chaboyer}, B. 2012, \apj, 757, 42

\bibitem[{{Fischer} \& {Valenti}(2005)}]{Fischer2005}
{Fischer}, D.~A., \& {Valenti}, J. 2005, \apj, 622, 1102

\bibitem[{{Fressin} {et~al.}(2013){Fressin}, {Torres}, {Charbonneau}, {Bryson},
  {Christiansen}, {Dressing}, {Jenkins}, {Walkowicz}, \&
  {Batalha}}]{Fressin2013}
{Fressin}, F., {et~al.} 2013, \apj, 766, 81

\bibitem[{{Girardi} {et~al.}(2005){Girardi}, {Groenewegen}, {Hatziminaoglou},
  \& {da Costa}}]{Girardi2005}
{Girardi}, L., {Groenewegen}, M.~A.~T., {Hatziminaoglou}, E., \& {da Costa}, L.
  2005, \aap, 436, 895

\bibitem[{{Gonzalez}(1997)}]{Gonzalez1997}
{Gonzalez}, G. 1997, \mnras, 285, 403

\bibitem[{{Gould} {et~al.}(2006){Gould}, {Dorsher}, {Gaudi}, \&
  {Udalski}}]{Gould2006}
{Gould}, A., {Dorsher}, S., {Gaudi}, B.~S., \& {Udalski}, A. 2006, actaa, 56,
  1

\bibitem[{{Henry} {et~al.}(1997){Henry}, {Ianna}, {Kirkpatrick}, \&
  {Jahreiss}}]{Henry1997}
{Henry}, T.~J., {Ianna}, P.~A., {Kirkpatrick}, J.~D., \& {Jahreiss}, H. 1997,
  \aj, 114, 388

\bibitem[{{Hinkel} {et~al.}(2014){Hinkel}, {Timmes}, {Young}, {Pagano}, \&
  {Turnbull}}]{Hinkel2014}
{Hinkel}, N.~R., {Timmes}, F.~X., {Young}, P.~A., {Pagano}, M.~D., \&
  {Turnbull}, M.~C. 2014, \aj, 148, 54

\bibitem[{{Horch} {et~al.}(2014){Horch}, {Howell}, {Everett}, \&
  {Ciardi}}]{Horch2014}
{Horch}, E.~P., {Howell}, S.~B., {Everett}, M.~E., \& {Ciardi}, D.~R. 2014,
  \apj, 795, 60

\bibitem[{{Howard} {et~al.}(2012){Howard}, {Marcy}, {Bryson}, {Jenkins},
  {Rowe}, {Batalha}, {Borucki}, {Koch}, {Dunham}, {Gautier}, {Van Cleve},
  {Cochran}, {Latham}, {Lissauer}, {Torres}, {Brown}, {Gilliland}, {Buchhave},
  {Caldwell}, {Christensen-Dalsgaard}, {Ciardi}, {Fressin}, {Haas}, {Howell},
  {Kjeldsen}, {Seager}, {Rogers}, {Sasselov}, {Steffen}, {Basri},
  {Charbonneau}, {Christiansen}, {Clarke}, {Dupree}, {Fabrycky}, {Fischer},
  {Ford}, {Fortney}, {Tarter}, {Girouard}, {Holman}, {Johnson}, {Klaus},
  {Machalek}, {Moorhead}, {Morehead}, {Ragozzine}, {Tenenbaum}, {Twicken},
  {Quinn}, {Isaacson}, {Shporer}, {Lucas}, {Walkowicz}, {Welsh}, {Boss},
  {Devore}, {Gould}, {Smith}, {Morris}, {Prsa}, {Morton}, {Still}, {Thompson},
  {Mullally}, {Endl}, \& {MacQueen}}]{Howard2012}
{Howard}, A.~W., {et~al.} 2012, \apjs, 201, 15

\bibitem[{{Howell} {et~al.}(2014){Howell}, {Sobeck}, {Haas}, {Still},
  {Barclay}, {Mullally}, {Troeltzsch}, {Aigrain}, {Bryson}, {Caldwell},
  {Chaplin}, {Cochran}, {Huber}, {Marcy}, {Miglio}, {Najita}, {Smith},
  {Twicken}, \& {Fortney}}]{Howell2014}
{Howell}, S.~B., {et~al.} 2014, \pasp, 126, 398

\bibitem[{{Huber} {et~al.}(2014){Huber}, {Silva Aguirre}, {Matthews},
  {Pinsonneault}, {Gaidos}, {Garc{\'{\i}}a}, {Hekker}, {Mathur}, {Mosser},
  {Torres}, {Bastien}, {Basu}, {Bedding}, {Chaplin}, {Demory}, {Fleming},
  {Guo}, {Mann}, {Rowe}, {Serenelli}, {Smith}, \& {Stello}}]{Huber2014}
{Huber}, D., {et~al.} 2014, \apjs, 211, 2

\bibitem[{{Jenkins} {et~al.}(2010{\natexlab{a}}){Jenkins}, {Chandrasekaran},
  {McCauliff}, {Caldwell}, {Tenenbaum}, {Li}, {Klaus}, {Cote}, \&
  {Middour}}]{Jenkins2010}
{Jenkins}, J.~M., {et~al.} 2010{\natexlab{a}}, in Society of Photo-Optical
  Instrumentation Engineers (SPIE) Conference Series, Vol. 7740, Society of
  Photo-Optical Instrumentation Engineers (SPIE) Conference Series

\bibitem[{{Jenkins} {et~al.}(2010{\natexlab{b}}){Jenkins}, {Chandrasekaran},
  {McCauliff}, {Caldwell}, {Tenenbaum}, {Li}, {Klaus}, {Cote}, \&
  {Middour}}]{Jenkins2010b}
{Jenkins}, J.~M., {et~al.} 2010{\natexlab{b}}, in Society of Photo-Optical
  Instrumentation Engineers (SPIE) Conference Series, Vol. 7740, Society of
  Photo-Optical Instrumentation Engineers (SPIE) Conference Series

\bibitem[{{Johnson} {et~al.}(2010){Johnson}, {Aller}, {Howard}, \&
  {Crepp}}]{Johnson2010}
{Johnson}, J.~A., {Aller}, K.~M., {Howard}, A.~W., \& {Crepp}, J.~R. 2010,
  \pasp, 122, 905

\bibitem[{{Kopparapu} {et~al.}(2013){Kopparapu}, {Ramirez}, {Kasting}, {Eymet},
  {Robinson}, {Mahadevan}, {Terrien}, {Domagal-Goldman}, {Meadows}, \&
  {Deshpande}}]{Kopparapu2013}
{Kopparapu}, R.~K., {et~al.} 2013, \apj, 765, 131

\bibitem[{{Kraus} \& {Hillenbrand}(2007)}]{Kraus2007}
{Kraus}, A.~L., \& {Hillenbrand}, L.~A. 2007, \aj, 134, 2340

\bibitem[{{Lissauer}(1993)}]{Lissauer1993}
{Lissauer}, J.~J. 1993, \araa, 31, 129

\bibitem[{{Lissauer} {et~al.}(2011){Lissauer}, {Ragozzine}, {Fabrycky},
  {Steffen}, {Ford}, {Jenkins}, {Shporer}, {Holman}, {Rowe}, {Quintana},
  {Batalha}, {Borucki}, {Bryson}, {Caldwell}, {Carter}, {Ciardi}, {Dunham},
  {Fortney}, {Gautier}, {Howell}, {Koch}, {Latham}, {Marcy}, {Morehead}, \&
  {Sasselov}}]{Lissauer2011}
{Lissauer}, J.~J., {et~al.} 2011, \apjs, 197, 8

\bibitem[{{Marcy} {et~al.}(2005){Marcy}, {Butler}, {Fischer}, {Vogt}, {Wright},
  {Tinney}, \& {Jones}}]{Marcy2005}
{Marcy}, G., {Butler}, R.~P., {Fischer}, D., {Vogt}, S., {Wright}, J.~T.,
  {Tinney}, C.~G., \& {Jones}, H.~R.~A. 2005, Progress of Theoretical Physics
  Supplement, 158, 24

\bibitem[{{Marcy} \& {Butler}(1996)}]{Marcy1996}
{Marcy}, G.~W., \& {Butler}, R.~P. 1996, \apjl, 464, L147

\bibitem[{{Mayor} \& {Queloz}(1995)}]{Mayor1995}
{Mayor}, M., \& {Queloz}, D. 1995, \nat, 378, 355

\bibitem[{{Mayor} {et~al.}(2011){Mayor}, {Marmier}, {Lovis}, {Udry},
  {S{\'e}gransan}, {Pepe}, {Benz}, {Bertaux}, {Bouchy}, {Dumusque}, {Lo Curto},
  {Mordasini}, {Queloz}, \& {Santos}}]{Mayor2011}
{Mayor}, M., {et~al.} 2011, ArXiv e-prints

\bibitem[{{Mortier} {et~al.}(2012){Mortier}, {Santos}, {Sozzetti}, {Mayor},
  {Latham}, {Bonfils}, \& {Udry}}]{Mortier2012}
{Mortier}, A., {Santos}, N.~C., {Sozzetti}, A., {Mayor}, M., {Latham}, D.,
  {Bonfils}, X., \& {Udry}, S. 2012, \aap, 543, A45

\bibitem[{{Morton} \& {Johnson}(2011)}]{Morton2011}
{Morton}, T.~D., \& {Johnson}, J.~A. 2011, \apj, 738, 170

\bibitem[{{Raghavan} {et~al.}(2010){Raghavan}, {McAlister}, {Henry}, {Latham},
  {Marcy}, {Mason}, {Gies}, {White}, \& {ten Brummelaar}}]{Raghavan2010}
{Raghavan}, D., {et~al.} 2010, \apjs, 190, 1

\bibitem[{{Ricker} {et~al.}(2014){Ricker}, {Winn}, {Vanderspek}, {Latham},
  {Bakos}, {Bean}, {Berta-Thompson}, {Brown}, {Buchhave}, {Butler}, {Butler},
  {Chaplin}, {Charbonneau}, {Christensen-Dalsgaard}, {Clampin}, {Deming},
  {Doty}, {De Lee}, {Dressing}, {Dunham}, {Endl}, {Fressin}, {Ge}, {Henning},
  {Holman}, {Howard}, {Ida}, {Jenkins}, {Jernigan}, {Johnson}, {Kaltenegger},
  {Kawai}, {Kjeldsen}, {Laughlin}, {Levine}, {Lin}, {Lissauer}, {MacQueen},
  {Marcy}, {McCullough}, {Morton}, {Narita}, {Paegert}, {Palle}, {Pepe},
  {Pepper}, {Quirrenbach}, {Rinehart}, {Sasselov}, {Sato}, {Seager},
  {Sozzetti}, {Stassun}, {Sullivan}, {Szentgyorgyi}, {Torres}, {Udry}, \&
  {Villasenor}}]{Ricker2014}
{Ricker}, G.~R., {et~al.} 2014, in Society of Photo-Optical Instrumentation
  Engineers (SPIE) Conference Series, Vol. 9143, Society of Photo-Optical
  Instrumentation Engineers (SPIE) Conference Series, 20

\bibitem[{{Santerne} {et~al.}(2012){Santerne}, {D{\'{\i}}az}, {Moutou},
  {Bouchy}, {H{\'e}brard}, {Almenara}, {Bonomo}, {Deleuil}, \&
  {Santos}}]{Santerne2012}
{Santerne}, A., {et~al.} 2012, \aap, 545, A76

\bibitem[{{Santos} {et~al.}(2001){Santos}, {Israelian}, \&
  {Mayor}}]{Santos2001}
{Santos}, N.~C., {Israelian}, G., \& {Mayor}, M. 2001, \aap, 373, 1019

\bibitem[{{Santos} {et~al.}(2004){Santos}, {Israelian}, \&
  {Mayor}}]{Santos2004}
---. 2004, \aap, 415, 1153

\bibitem[{{Schlaufman} \& {Winn}(2013)}]{Schlaufman2013}
{Schlaufman}, K.~C., \& {Winn}, J.~N. 2013, \apj, 772, 143

\bibitem[{{Seager}(2013)}]{Seager2013}
{Seager}, S. 2013, Science, 340, 577

\bibitem[{{Sousa} {et~al.}(2011){Sousa}, {Santos}, {Israelian}, {Mayor}, \&
  {Udry}}]{Sousa2011}
{Sousa}, S.~G., {Santos}, N.~C., {Israelian}, G., {Mayor}, M., \& {Udry}, S.
  2011, \aap, 533, A141

\bibitem[{{Udry} \& {Santos}(2007)}]{Udry2007}
{Udry}, S., \& {Santos}, N.~C. 2007, \araa, 45, 397

\bibitem[{{Valenti} \& {Fischer}(2005)}]{Valenti2005}
{Valenti}, J.~A., \& {Fischer}, D.~A. 2005, \apjs, 159, 141

\bibitem[{{Wang} \& {Fischer}(2013)}]{Wang2013b}
{Wang}, J., \& {Fischer}, D.~A. 2013, ArXiv e-prints

\bibitem[{{Wang} {et~al.}(2014{\natexlab{a}}){Wang}, {Fischer}, {Xie}, \&
  {Ciardi}}]{Wang2014b}
{Wang}, J., {Fischer}, D.~A., {Xie}, J.-W., \& {Ciardi}, D.~R.
  2014{\natexlab{a}}, \apj, 791, 111

\bibitem[{{Wang} {et~al.}(2014{\natexlab{b}}){Wang}, {Xie}, {Barclay}, \&
  {Fischer}}]{Wang2014a}
{Wang}, J., {Xie}, J.-W., {Barclay}, T., \& {Fischer}, D.~A.
  2014{\natexlab{b}}, \apj, 783, 4

\bibitem[{{Wright} {et~al.}(2012){Wright}, {Marcy}, {Howard}, {Johnson},
  {Morton}, \& {Fischer}}]{Wright2012}
{Wright}, J.~T., {Marcy}, G.~W., {Howard}, A.~W., {Johnson}, J.~A., {Morton},
  T.~D., \& {Fischer}, D.~A. 2012, \apj, 753, 160

\bibitem[{{Wright} {et~al.}(2011){Wright}, {Fakhouri}, {Marcy}, {Han}, {Feng},
  {Johnson}, {Howard}, {Fischer}, {Valenti}, {Anderson}, \&
  {Piskunov}}]{Wright2011}
{Wright}, J.~T., {et~al.} 2011, \pasp, 123, 412

\bibitem[{{Wu} \& {Murray}(2003)}]{Wu2003}
{Wu}, Y., \& {Murray}, N. 2003, \apj, 589, 605

\end{thebibliography}

%

\clearpage

%
%
%


\clearpage

\begin{table}
\caption{Current knowledge of parameters in Equation \ref{eq:f_hj_sgr} and \ref{eq:f_hj_mr} and the differences of stellar properties and stellar multiplicity between the Doppler sample and the Kepler sample. 
\label{tab:current}}
\begin{tabular}{lcc}
&Kepler&Doppler\\
$f$&\checkmark&\checkmark\\
$\xi_{\rm{[Fe/H]}}$&\multicolumn{2}{c}{\halfcheck}\\
$\xi_{\rm{misidentified}}$&\multicolumn{2}{c}{\halfcheck}\\
FPR&\multicolumn{2}{c}{\halfcheck}\\
SGR& \halfcheck&\checkmark\\
MR$_{\rm{MS}}$& \halfcheck& \halfcheck\\
MR$_{\rm{SG}}$& \halfcheck& \halfcheck\\

\end{tabular}
\tablecomments{$f$ - measured HJ occurrence rate; $\xi_{\rm{[Fe/H]}}$ - correcting factor for the metallicity difference; $\xi_{\rm{misidentified}}$ - correcting factor for misidentified HJs due to photometric dilution and subgiant contamination; FPR - false positive rate of Kepler planet candidates; SGR - fraction of subgiants; MR$_{\rm{MS}}$ - stellar multiplicity rate for main sequence stars; MR$_{\rm{SG}}$ - stellar multiplicity rate for subgiants. }
\end{table}

\end{document}